\RequirePackage{lineno}
\documentclass[aps,prl,twocolumn,showpacs,preprintnumbers,superscriptaddress,floatfix]{revtex4}

\usepackage{graphicx}  
\usepackage{dcolumn}   
\usepackage{bm}        
\usepackage{amssymb}   
\usepackage{setspace}
\usepackage{amsmath, amssymb, setspace}
\usepackage{array}
\usepackage{booktabs}
\usepackage{caption}
\usepackage{mathrsfs}
\usepackage{indentfirst}
\usepackage{slashed}
\usepackage{float}
\usepackage{lmodern}
\usepackage{hyperref}
\usepackage{xcolor}
\usepackage{multirow}
\usepackage{epstopdf}
\usepackage{ulem}

\begin{document}

\newcommand{\VT}{\textcolor{black}}


\title{Search for Trilepton Nucleon Decay via $p \rightarrow  e^+ \nu \nu$ and $p \rightarrow \mu^+ \nu \nu$ in the Super-Kamiokande Experiment}

\newcommand{\AFFicrr}{\affiliation{Kamioka Observatory, Institute for Cosmic Ray Research, University of Tokyo, Kamioka, Gifu 506-1205, Japan}}
\newcommand{\AFFkashiwa}{\affiliation{Research Center for Cosmic Neutrinos, Institute for Cosmic Ray Research, University of Tokyo, Kashiwa, Chiba 277-8582, Japan}}
\newcommand{\AFFipmu}{\affiliation{Kavli Institute for the Physics and
Mathematics of the Universe (WPI), Todai Institutes for Advanced Study,
University of Tokyo, Kashiwa, Chiba 277-8582, Japan }}
\newcommand{\AFFmad}{\affiliation{Department of Theoretical Physics, University Autonoma Madrid, 28049 Madrid, Spain}}
\newcommand{\AFFubc}{\affiliation{Department of Physics and Astronomy, University of British Columbia, Vancouver, BC, V6T1Z4, Canada}}
\newcommand{\AFFbu}{\affiliation{Department of Physics, Boston University, Boston, MA 02215, USA}}
\newcommand{\AFFbnl}{\affiliation{Physics Department, Brookhaven National Laboratory, Upton, NY 11973, USA}}
\newcommand{\AFFuci}{\affiliation{Department of Physics and Astronomy, University of California, Irvine, Irvine, CA 92697-4575, USA }}
\newcommand{\AFFcsu}{\affiliation{Department of Physics, California State University, Dominguez Hills, Carson, CA 90747, USA}}
\newcommand{\AFFcnm}{\affiliation{Department of Physics, Chonnam National University, Kwangju 500-757, Korea}}
\newcommand{\AFFduke}{\affiliation{Department of Physics, Duke University, Durham NC 27708, USA}}
\newcommand{\AFFfukuoka}{\affiliation{Junior College, Fukuoka Institute of Technology, Fukuoka, Fukuoka 811-0295, Japan}}
\newcommand{\AFFgmu}{\affiliation{Department of Physics, George Mason University, Fairfax, VA 22030, USA }}
\newcommand{\AFFgifu}{\affiliation{Department of Physics, Gifu University, Gifu, Gifu 501-1193, Japan}}
\newcommand{\AFFgist}{\affiliation{GIST College, Gwangju Institute of Science and Technology, Gwangju 500-712, Korea}}
\newcommand{\AFFuh}{\affiliation{Department of Physics and Astronomy, University of Hawaii, Honolulu, HI 96822, USA}}
\newcommand{\AFFkanagawa}{\affiliation{Physics Division, Department of Engineering, Kanagawa University, Kanagawa, Yokohama 221-8686, Japan}}
\newcommand{\AFFkek}{\affiliation{High Energy Accelerator Research Organization (KEK), Tsukuba, Ibaraki 305-0801, Japan }}
\newcommand{\AFFkobe}{\affiliation{Department of Physics, Kobe University, Kobe, Hyogo 657-8501, Japan}}
\newcommand{\AFFkyoto}{\affiliation{Department of Physics, Kyoto University, Kyoto, Kyoto 606-8502, Japan}}
\newcommand{\AFFumd}{\affiliation{Department of Physics, University of Maryland, College Park, MD 20742, USA }}
\newcommand{\AFFmit}{\affiliation{Department of Physics, Massachusetts Institute of Technology, Cambridge, MA 02139, USA}}
\newcommand{\AFFmiyagi}{\affiliation{Department of Physics, Miyagi University of Education, Sendai, Miyagi 980-0845, Japan}}
\newcommand{\AFFnagoya}{\affiliation{Solar Terrestrial Environment Laboratory, Nagoya University, Nagoya, Aichi 464-8602, Japan}}
\newcommand{\AFFpol}{\affiliation{National Centre For Nuclear Research, 00-681 Warsaw, Poland}}
\newcommand{\AFFsuny}{\affiliation{Department of Physics and Astronomy, State University of New York at Stony Brook, NY 11794-3800, USA}}
\newcommand{\AFFniigata}{\affiliation{Department of Physics, Niigata University, Niigata, Niigata 950-2181, Japan }}
\newcommand{\AFFokayama}{\affiliation{Department of Physics, Okayama University, Okayama, Okayama 700-8530, Japan }}
\newcommand{\AFFosaka}{\affiliation{Department of Physics, Osaka University, Toyonaka, Osaka 560-0043, Japan}}
\newcommand{\AFFregina}{\affiliation{Department of Physics, University of Regina, 3737 Wascana Parkway, Regina, SK, S4SOA2, Canada}}
\newcommand{\AFFseoul}{\affiliation{Department of Physics, Seoul National University, Seoul 151-742, Korea}}
\newcommand{\AFFshizuokasc}{\affiliation{Department of Informatics in
Social Welfare, Shizuoka University of Welfare, Yaizu, Shizuoka, 425-8611, Japan}}
\newcommand{\AFFskk}{\affiliation{Department of Physics, Sungkyunkwan University, Suwon 440-746, Korea}}
\newcommand{\AFFtohoku}{\affiliation{Research Center for Neutrino Science, Tohoku University, Sendai, Miyagi 980-8578, Japan}}
\newcommand{\AFFtokyo}{\affiliation{The University of Tokyo, Bunkyo, Tokyo 113-0033, Japan }}
\newcommand{\AFFToronto}{\affiliation{Department of Physics, University of Toronto, 60 St., Toronto, Ontario, M5S1A7, Canada }}
\newcommand{\AFFtriumf}{\affiliation{TRIUMF, 4004 Wesbrook Mall, Vancouver, BC, V6T2A3, Canada }}
\newcommand{\AFFtokai}{\affiliation{Department of Physics, Tokai University, Hiratsuka, Kanagawa 259-1292, Japan}}
\newcommand{\AFFtit}{\affiliation{Department of Physics, Tokyo Institute
for Technology, Meguro, Tokyo 152-8551, Japan }}
\newcommand{\AFFtsinghua}{\affiliation{Department of Engineering Physics, Tsinghua University, Beijing, 100084, China}}
\newcommand{\AFFwarsaw}{\affiliation{Institute of Experimental Physics, Warsaw University, 00-681 Warsaw, Poland }}
\newcommand{\AFFuw}{\affiliation{Department of Physics, University of Washington, Seattle, WA 98195-1560, USA}}

\AFFicrr
\AFFkashiwa
\AFFmad
\AFFbu
\AFFubc
\AFFbnl
\AFFuci
\AFFcsu
\AFFcnm
\AFFduke
\AFFfukuoka
\AFFgifu
\AFFgist
\AFFuh
\AFFkek
\AFFkobe
\AFFkyoto
\AFFmiyagi
\AFFnagoya
\AFFsuny
\AFFokayama
\AFFosaka
\AFFregina
\AFFseoul
\AFFshizuokasc
\AFFskk
\AFFtokai
\AFFtokyo
\AFFipmu
\AFFToronto
\AFFtriumf
\AFFtsinghua
\AFFuw

\author{V.~Takhistov} 
\AFFuci

\author{K.~Abe}
\AFFicrr
\AFFipmu
\author{Y.~Haga}
\AFFicrr
\author{Y.~Hayato}
\AFFicrr
\AFFipmu
\author{M.~Ikeda}
\AFFicrr
\author{K.~Iyogi}
\AFFicrr 
\author{J.~Kameda}
\author{Y.~Kishimoto}
\author{M.~Miura} 
\author{S.~Moriyama} 
\author{M.~Nakahata}
\AFFicrr
\AFFipmu 
% ...

\collaboration{The Super-Kamiokande Collaboration}
\noaffiliation

\date{\today}

\begin{abstract}
The trilepton nucleon decay modes $p \rightarrow  e^+ \nu \nu$ and $p \rightarrow \mu^+ \nu \nu$ 
violate $|\Delta (B - L)|$ by two units. 
Using data from a 273.4 kiloton year exposure of Super-Kamiokande a search for these decays
yields a fit consistent with no signal. Accordingly, lower limits on the partial lifetimes of
$\tau_{p \rightarrow  e^+ \nu \nu} > 1.7 \times 10^{32}$ years and 
$\tau_{p \rightarrow \mu^+ \nu \nu} > 2.2 \times 10^{32}$ years at a $90 \% $ confidence level are obtained.
These limits can constrain Grand Unified Theories which allow for such processes.
\end{abstract}

\pacs{}
\pacs{12.10.Dm,13.30.-a,11.30.Fs,12.60.Jv,14.20.Dh,29.40.Ka} 


\maketitle

There is strong theoretical motivation for a Grand Unified Theory (GUT) \cite{Georgi:1974sy,Fritzsch:1974nn}
as an underlying description of nature.~Unification of the running couplings, charge 
quantization, as well as other hints point to the Standard Model (SM) being an incomplete
theory.~Though the GUT energy scale is inaccessible to accelerator experiments
a signature prediction of these theories is an unstable proton with 
lifetimes that can be probed by large underground experiments. 
Observation of proton decay would constitute strong evidence for physics beyond the SM,
 and non-observation imposes stringent constraints on GUT models.

One of the simplest unification scenarios{\color{red},} 
based on minimal SU(5), has been decisively ruled
out by limits on $p \rightarrow e^+ \pi^0$ \cite{McGrew:1999nd, Hirata:1989kn, Shiozawa:1998si}. 
On the other hand, models based on minimal supersymmetric (SUSY) extensions are strongly 
constrained by bounds from $p \rightarrow \bar{\nu} K^+$ \cite{Kobayashi:2005pe}, and with
signs of SUSY unobserved at the Large Hadron Collider (LHC) \cite{Chatrchyan:2013sza, Aad:2011hh}, 
there is reinvigorated interest in other approaches and possible signatures.~A popular scenario may be found 
in a left-right symmetric partial unification of
 Pati and Salam (PS) \cite{Pati:1974yy} and its embedding into SO(10),
providing a natural right-handed neutrino candidate 
and unifying quarks and leptons.
In the scheme of Ref.~\cite{Pati:1983zp, Pati:1983jk},
trilepton modes such as $p \rightarrow  e^+ \nu \nu$ and $p \rightarrow \mu^+ \nu \nu$
could become significant.
This work describes searches for these modes.  Their
observation, coupled with non-observation of $p \rightarrow e^+ \pi^0$, may allow for differentiation
between PS and its SO(10) embedding \cite{Pati:1983jk}.~Violating
 baryon and lepton number by two units ($|\Delta (B - L)| = 2)$, unusual
for standard decay channels, may lead 
to favorable implications for baryogenesis \cite{Gu:2011pf}. 
Interestingly, these trilepton proton decay
 modes were offered as an explanation \cite{Mann:1992ue, O'Donnell:1993db}
of the atmospheric neutrino flavor 
``anomaly" \cite{BeckerSzendy:1992hq,Fukuda:1994mc}
 before neutrino oscillations were established \cite{Fukuda:1998mi}.

In this analysis,~the data collected at Super-Kamiokande (SK) during the data taking periods 
of SK-I (May 1996-Jul 2001, 1489.2 live days), SK-II (Jan 2003-Oct 2005, 798.6 live days), 
SK-III (Sept 2006-Aug 2008, 518.1 live days) and the ongoing SK-IV experiment 
(Sept 2008-Oct 2013, 1632.3 live days), corresponding to a combined
exposure of 273.4 kton $\cdot$ years, is analyzed. The 50 kiloton SK water Cherenkov
detector (22.5 kton fiducial volume) is located beneath a one-km rock overburden 
(2700m water equivalent) in the Kamioka mine in Japan. 
Details of the detector design and performance in each SK period,
as well as calibration, data reduction and simulation 
information can be found elsewhere~\cite{Fukuda:2002uc,Abe:2013gga}.
This analysis considers only events in which 
all observed Cherenkov light was fully contained within the inner detector.

The trilepton decay modes $p \rightarrow  e^+ \nu \nu$  and $p \rightarrow \mu^+ \nu \nu$ 
are the first three-body nucleon decay searches undertaken by SK. 
Since the neutrinos cannot be observed,
the only signature is the appearance of a charged lepton, $e^+$ or $\mu^+$.
Accordingly, the invariant mass of the decay nucleon cannot be reconstructed.
Unlike two-body decays, where each final-state particle
carries away about half of the nucleon rest mass energy,
in these three-body decays the charged lepton has a broad energy distribution,
whose mean is 313 MeV for the decay of a free proton.  
Thus, atmospheric neutrino interactions dominate the lepton energy spectra and
require a search for the proton decay signal superimposed on a substantial background.  
Limits on these modes from the 
IMB-3 \cite{McGrew:1999nd} and Fr\'ejus \cite{Berger:1991fa} experiments, 
$1.7 \times 10^{31}$  and $2.1 \times 10^{31}$ years, were obtained 
via simple counting techniques.  
In contrast, we employ energy spectrum fits.  This technique is particularly well suited
to three-body searches with large backgrounds as it takes full
advantage of the signal and background spectral information.

The detection efficiency for nucleon decays in water is estimated
from Monte Carlo (MC) simulations in which
all protons within the H$_2$O molecule are assumed to decay with equal probability. 
Signal events are obtained by generating final state particles from the proton's decay
with energy and momentum uniformly distributed within the phase space.~Conservation 
of kinematic variables constrain the processes to produce viable particle spectra.
Specifics of the decay dynamics, 
which are model dependent but are not taken into account here,
can play a role in determining the energy 
distributions of the resulting particles in three body decays.
\VT{The assumption of a flat phase space, as employed within this analysis, was validated by comparing the final state charged lepton spectrum generated with a flat phase space to the spectrum originating from the three-body phase space of muon decay (reaction), as recently proposed  \cite{Chen:2014ifa} to account for decays encompassing a broad range of models. We have confirmed that adopting a non-flat phase space does not significantly alter the results of the analysis, because the charged lepton spectra do not have sufficiently different shapes (even for the decay of a free proton, which is minimally smeared).
Thus, we conclude, that employing} flat phase space in the signal \
simulation\VT{, which has been previously assumed in other similar searches
\cite{McGrew:1999nd,Berger:1991fa} without much justification, is warranted.}

In the signal simulation, the effects of Fermi momentum and the nuclear binding energy 
as well as nucleon-nucleon correlated decays 
are taken into account~\cite{Nishino:2012ipa, Regis:2012sn}.
Fermi momentum distributions are simulated using a spectral function
fit to $^{12}$C electron scattering data~\cite{Nakamura:1976mb}.
Considering only events generated within the fiducial volume (FV) of the detector, 
the signal MC consists of roughly 4000 events for each of the SK data periods.

Atmospheric neutrino background interactions are generated 
using the flux of Honda \textit{et. al.}~\cite{Honda:2006qj}
and the NEUT simulation package~\cite{Hayato:2002sd},
which uses a relativistic Fermi gas model.
The SK detector simulation~\cite{Abe:2013gga} is based on the GEANT-3 \cite{Brun:1994aa} package. 
Background MC corresponding to a 500 year exposure of the detector is generated for each SK period.

The following event selection criteria are applied to the fully-contained data:~(A) a single Cherenkov ring is present, 
(B) the ring is showering (electron-like) for $p \rightarrow  e^+ \nu \nu$
 and non-showering (muon-like) for $p \rightarrow \mu^+ \nu \nu$,
(C) there are zero decay electrons for $p \rightarrow  e^+ \nu \nu$
 and one decay electron for $p \rightarrow \mu^+ \nu \nu$,
(D) the reconstructed momentum lies in the range 100 MeV/$c$ $ \le p_e \le $ 1000 MeV/$c$
 for $p \rightarrow  e^+ \nu \nu$ and in the range 200 MeV/$c$ $ \le p_\mu \le $ 1000 MeV/$c$
 for $p \rightarrow \mu^+ \nu \nu$. Reconstruction details may be found in Ref. \cite{Shiozawa:1999sd}.
The signal detection efficiency is defined as the fraction of events passing these selection criteria
compared to the total number of events generated within the true fiducial volume (see Table \ref{tab:results}).
The increase in efficiency seen in SK-IV for the $p \rightarrow \mu^+ \nu \nu$ mode is caused by a 20\% improvement in
the detection of muon decay electrons after an upgrade of the detector electronics for this period~\cite{Abe:2013gga}.

\begin{table*}[htb]
\begin{center}
\caption{Best fit parameter values, signal detection efficiency 
         for each SK running period, 90\%~C.L. value of $\beta$ parameter,
         allowed number of nucleon decay events in the full 273.4~kiloton $\cdot$ year
         exposure (SK-I: 91.7, SK-II: 49.2, SK-III: 31.9, SK-IV: 100.5) and a partial
         lifetime limit for each decay mode at 90\%
           C.L.}  \begin{tabular}{lcccccc} \hline \hline
  Decay mode   &  Best fit values   & Signal efficiency & $\beta_{{\mathrm{90CL}}}$ &  Signal events at 90\%~C.L. &  $\tau/{\mathcal{B}}$ \\ 
               &  $(\alpha,\beta)$  &  for SK-I, -II, -III, -IV (\%)   & & ($N_{{\mathrm{90CL}}}$) & ($\times10^{32}$~yrs)\\
               &  	  & (efficiency uncertainty)   & 	& 	 &	\\
  \hline
   $p \rightarrow  e^+ \nu \nu$		&   (1.05, 0.03)	  &  88.8, 88.0, 89.2, 87.8 	 					& 0.06 & 459 & 1.7\\
  											& 		  			  &  ($\pm$0.5, $\pm$0.5, $\pm$0.5, $\pm$0.5)  & 	 	&  		 &  \\			
\hline		
   $p \rightarrow \mu^+ \nu \nu$	 &   (0.99, 0.02)   &  64.4, 65.0, 67.0, 78.4						& 0.05 & 286 & 2.2\\
  											& 		  			  &  ($\pm$0.7, $\pm$0.7, $\pm$0.7, $\pm$0.6)   & 	 	&  		 &  \\			
  \hline 
  \hline 
  \end{tabular} 
   \label{tab:results}
\end{center}
\end{table*}

\begin{table*}[htb]
\begin{center}
\caption{Systematic errors of the nucleon decay spectrum fits, with $1\sigma$
         uncertainties and resulting fit pull terms. Errors specific to signal and background are denoted by S and B, while those that are common to both by SB.}  
  \begin{tabular}{lccccc} \hline \hline
   		Decay mode			 & ~ & $p \rightarrow e^+ \nu \nu$ &  $p \rightarrow \mu^+ \nu \nu$ &\\
  \hline
    Systematic error  & 	1-$\sigma$ uncertainty (\%) & Fit pull ($\sigma$)	 & Fit pull ($\sigma$)  & \\
  \hline
    Final state interactions (FSI)										& 10  				    &  ~0.08 			 					  &  -0.55  						&~  B\\
    Flux normalization ($E_\nu < 1$ GeV)  							&~~25 	\footnote{Uncertainty linearly decreases with $\log{E_\nu}$ from 25\% (0.1 GeV) to 7\% (1 GeV).}	
			  &  -0.36		 						  &  -0.42  						& ~ B\\
    Flux normalization ($E_\nu > 1$ GeV)  							 &~~15  	\footnote{Uncertainty is 7\% up to 10 GeV, linearly increases with $\log{E_\nu}$ from 7\% (10 GeV) to 12\% (100 GeV) and then 20\% (1 TeV).}			  			
			     &  -0.86		  						  &  -0.90 						& ~ B\\
    $M_A$ in $\nu$ interactions							         				  &	10  		  	  		 &  ~0.32 		 						  &  ~0.48  							& ~ B\\
    Single meson cross-section in $\nu$ interactions			 &	10  			  		 &  -0.36  		   					  &  -0.16  						& ~ B\\
    Energy calibration of SK-I, -II, -III, -IV          					  & 1.1, 1.7, 2.7, 2.3  	  &  ~0.51, -1.01,  0.44, 0.39 	  	  & ~ -0.50, 0.06,  -0.16, 0.25	& ~ SB\\
    Fermi model comparison         									  &~~10   		\footnote{Comparison of spectral function and Fermi gas model.}	
				  &  -0.25  	  		 			       & ~0.02							& ~ S\\		
    Nucleon-nucleon correlated decay         						 &  100  				      &  -0.05		   					       &  ~0.01					&~ S\\
  \hline 
	\hline
  \end{tabular} 
   \label{tab:syserr}
\end{center}
\end{table*}

In the case of $p \rightarrow  e^+ \nu \nu$, the dominant (78\%) background 
after selection criteria are applied is due to  $\nu_e$
quasi-elastic charged current (CCQE) interactions.
The majority of the remaining background is due to $\nu_e$ and $\nu_\mu$ charged
current pion production (CC) as well as the 
all flavor's neutral current (NC) single pion production (12\% and 5\%, respectively).
There are minor contributions from other processes such as
coherent pion production (order of 1\%).~Similarly
 for the $p \rightarrow  \mu^+ \nu \nu$ mode, $\nu_\mu$
CCQE interactions dominate (80\%), with the largest remaining contribution
coming from CC single pion production (15\%). 
Additionally there are slight contributions from NC pion production,
CC coherent and multiple-pion production (around 1\% each).
Processes not mentioned here are negligible.

A spectrum fit is performed on the reconstructed charged
 lepton momentum distributions of selected candidates.
The foundation of the fit is a $\chi^2$ minimization
with systematic errors accounted for by quadratic penalties (``pull terms") as described 
in Ref.~\cite{Fogli:2002pt}.
The $\chi^2$ function is defined as

\begin{equation}
\begin{split}
%
%
& \chi^{2} = 2 \sum^{{\textrm{nbins}}}_{i=1} \Big( z_i - N^{{\textrm{obs}}}_{i} 
              +~N^{{\textrm{obs}}}_{i} \ln \frac{ N^{{\textrm{obs}}}_{i} }{ z_i } \Big)
             + \sum^{N_{{\textrm{syserr}}}}_{j=1} ( \frac{ \epsilon_{j} }{ \sigma_{j} } )^{2}  \\
& z_i = \alpha \cdot N^{{\textrm{back}}}_{i}( 1 + \sum^{N_{{\textrm{syserr}}}}_{j=1} f^{j}_{i} \frac{ \epsilon_{j} }{ \sigma_{j}}  ) 
+ \beta \cdot N^{{\textrm{sig}}}_{i}( 1 + \sum^{N_{{\textrm{syserr}}}}_{j=1} f^{j}_{i} \frac{ \epsilon_{j} }{ \sigma_{j}}  ), 
\end{split}
\label{eq:chi}
\end{equation}

\noindent where $i$ labels the analysis bins. 
The terms $N^{{\textrm{obs}}}_{i}$, $N^{{\textrm{sig}}}_{i}$, $N^{{\textrm{back}}}_{i}$
are the number of observed data, signal MC and background MC
events in bin $i$. The MC expectation in a bin is taken to be
$N^{{\textrm{exp}}}_{i} = \alpha \cdot N^{{\textrm{back}}}_{i} + \beta \cdot N^{{\textrm{sig}}}_{i}$, with $\alpha$ and $\beta$ denoting
the background (atmospheric neutrino) and signal (nucleon decay) normalizations. 
The $j^{th}$ systematic error is accounted for by the ``pull term", where $\epsilon_{j}$ 
is the fit error parameter and $f_{i}^{j}$ is the fractional change
in the MC expectation bin due to a 1 sigma uncertainty $\sigma_{j}$ of the error.~A
\VT{two-parameter} fit is performed to the parameters $\alpha$ and $\beta$,
with the point $(\alpha, \beta) = (1, 0)$ set to correspond to no signal hypothesis.
With signal spectrum normalized by area to the background prior to the fit, $\beta = 1$
corresponds to the amount of nucleon decay events equal to the quantity of
background MC after detector livetime normalization.
The parameter space of $(\alpha, \beta)$ is allowed to vary in
 the intervals of ($\alpha \in [0.8, 1.2], ~\beta \in [0.0, 0.2]$).
The $\chi^2$ of Eq.~(\ref{eq:chi}) is minimized with respect to $\epsilon_{j}$
 according to $ \partial \chi^2/  \partial \epsilon_{j} = 0$, yielding a set of 
equations which are solved iteratively, and the global minimum is defined
as the best fit. \VT{The confidence level intervals are later
derived from the $\chi^2$ minimization at each point in the
$(\alpha, \beta)$ plane after subtracting off this global minimum.
Namely, the CL limit is based on the constant $\Delta \chi^2$ criti-
cal value corresponding to the $90\%$ CL for a fit with one
degree of freedom, after profiling out $\beta$'s  dependence on $\alpha$
from the two-parameter fit.}

 Combining signal and background into each analysis MC expectation bin, 
as employed in a typical fit of this sort (see Ref.~\cite{Fogli:2002pt}),
is an approximate approach where systematic errors for signal as well as
 background are applied to every analysis bin which contains both.
In this analysis we employ a more accurate error treatment,
splitting signal and background (doubling the number of analysis bins) 
for the application of systematic errors 
and then recombining them during the $\chi^2$ minimization. 
A total of 72 momentum bins (18, 50-MeV/$c$ wide bins for each SK period)
are considered for $p \rightarrow  e^+ \nu \nu$, 
corresponding to 144 MC bins when the background and signal are separated. 
In the case of $p \rightarrow  \mu^+ \nu \nu$ a total of 64 momentum bins  
(16, 50-MeV/$c$ wide bins for each SK period) are used in the analysis,
corresponding to 128 MC bins with background and signal separated.

Systematic errors may be divided into several categories: 
background systematics, detector and reconstruction systematics, 
and signal systematics. Detector and reconstruction 
systematics are common to both signal and background.

This study starts by considering all 154 systematic uncertainties
which are taken into account in the standard SK neutrino oscillation
analysis \cite{Wendell:2010md}, along with two signal-specific systematic
effects related to correlated decays and Fermi momentum. 
In order to select which systematic uncertainties to include in the limit calculation,
 only error terms with at least
 one $|f_{i}^{j}|>0.05$ are used in the analysis. 
Loosening the selection to $|f_{i}^{j}|>0.01$ does not significantly
affect the analyses results but greatly increases the number of
 errors to be treated.~After selection, there are 11 systematic error
 terms for both $p \rightarrow  e^+ \nu \nu$ and $p \rightarrow  \mu^+ \nu \nu$. 
The main systematic contributions originate from energy calibration uncertainties 
(common error to both signal and background),
uncertainties related to the atmospheric neutrino flux, 
and uncertainties in the signal simulation. 
The complete list of errors, their uncertainties,
 and fitted pull terms can be found in Table~\ref{tab:syserr}. 
Errors specific to signal and background are denoted by S and B, respectively,
 while those that are common to both are denoted by SB.

\begin{figure*}[htb]
\minipage{0.5\textwidth}
\includegraphics[scale=0.45]{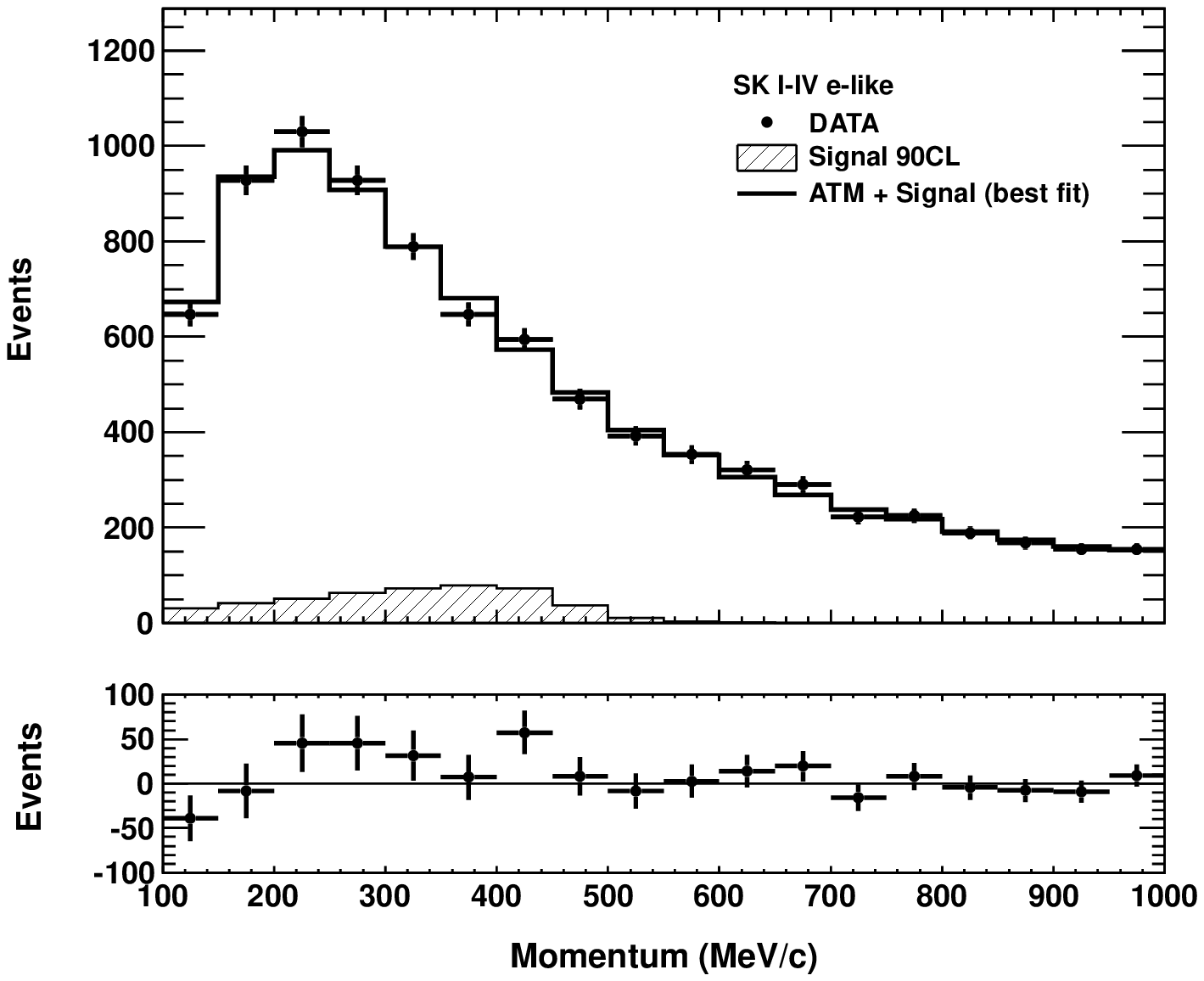}
\endminipage\hfill
\minipage{0.5\textwidth}%
\includegraphics[scale=0.45]{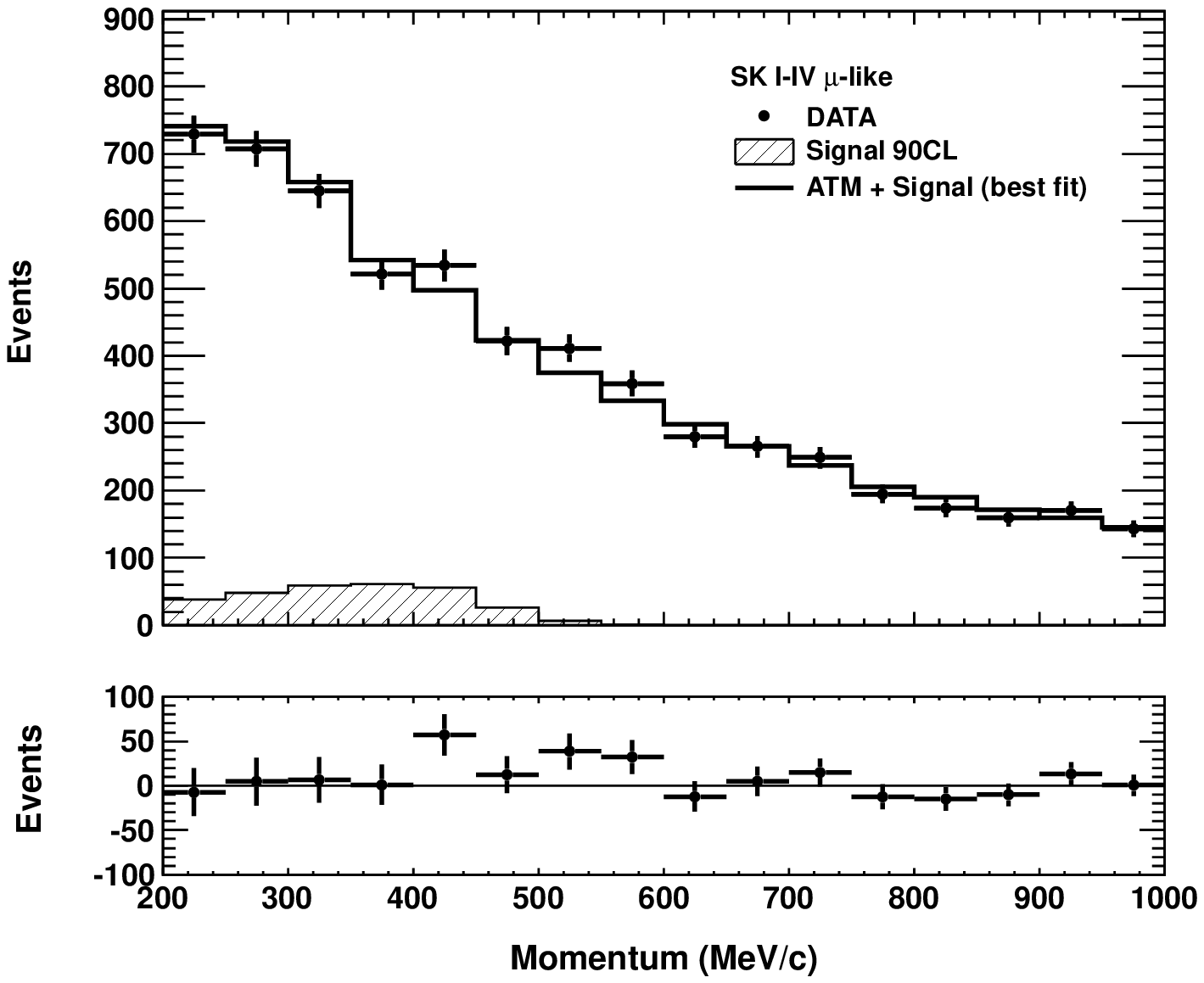}
\endminipage
\caption{Reconstructed momentum distribution for 273.4 kton $\cdot$ years of combined SK data (black dots), 
the best fit result for the atmospheric neutrino
background and signal Monte Carlo (solid line) as well as the 90\% confidence level allowed amount of nucleon decay (hatched histogram) for 
$p \rightarrow  e^+ \nu \nu$ (left) and $p \rightarrow \mu^+ \nu \nu$ (right). Residuals from data after background subtraction (bottom histograms).
}\label{fig:results_full}
\end{figure*}

Performing the fit allows us to obtain the overall background 
and signal normalizations $\alpha$ and $\beta$.
\VT{For  the mode $p \rightarrow  e^+ \nu \nu$  the data's best 
fit point is found to be $(\alpha, \beta) = (1.05, 0.03)$ with $\chi^2 = 65.6/ 70$ dof ,
while for $p \rightarrow  \mu^+ \nu \nu$ the result is $(\alpha, \beta)  = (0.99, 0.02)$
 with $\chi^2 = 66.1/ 62$ dof. 
The $\Delta \chi^2 (= \chi^2 - \chi^2_{\text{min}})$ values corresponding to no proton decay signal being present, are 1.5 and 0.5 for
$p \rightarrow  e^+ \nu \nu$ and $p \rightarrow  \mu^+ \nu \nu$ modes respectively.}
These outcomes are consistent with no signal present at 1 $\sigma$ level.
Extracting the 90\% confidence level allowed value of $\beta$ ($\beta_{90 \text{CL}}$) 
from the fit, which is found to be 0.06 for $p \rightarrow  e^+ \nu \nu$ and 
0.05 for $p \rightarrow  \mu^+ \nu \nu$ respectively, a 
lower lifetime limit on these decays can be set.
From $\beta_{90 \text{CL}}$ the amount of signal
allowed at the 90\% confidence level can be computed as 
$N_{90 \text{CL}} = \beta_{90 \text{CL}} \cdot N^{\text{signal}}$.
The partial lifetime limit for each decay mode is then calculated according to

\begin{equation}
	\tau_{90 \text{CL}} /\mathcal{B} ~=~\frac{\sum^{{\textrm{SK4}}}_{\text{sk} =\textrm{SK1}}\lambda_{\text{sk}} \cdot \epsilon_{\text{sk}} \cdot  N^{\text{nucleons}}}{N_{90\text{CL}}},
\end{equation}

\noindent where $\mathcal{B}$ represents the branching ratio of a process,
 $N^{\text{nucleons}}$ is the number of nucleons per kiloton of water ($3.3 \times 10^{32}$ protons),
$\epsilon_{\text{sk}}$ is the signal efficiency in each SK phase, 
$\lambda_{\text{sk}}$ is the corresponding exposure in kiloton $\cdot$ years,
 and $N_{90\text{CL}}$ is the amount of signal allowed at the 90\%
confidence level. The signal efficiency, number of decay sources,
 as well as the signal normalization 
values used for the lifetime calculation can be found in Table \ref{tab:results}. 
The fitted momentum spectra as well as residuals 
for both modes appear in Figure \ref{fig:results_full}.
Momentum spectra for the 273.4 kton $\cdot$ years of combined SK data (black dots),
the best-fit result for the atmospheric neutrino
background and signal Monte Carlo (solid line) as well as the amount of nucleon decay
allowed at the 90\% confidence level (hatched histogram) for 
$p \rightarrow  e^+ \nu \nu$ (left) and $p \rightarrow \mu^+ \nu \nu$ (right) are shown.
Residuals from data after background MC is subtracted are also depicted 
(bottom histograms).~From the analysis we set
 partial lifetime limits of $1.7 \times 10^{32}$ and $2.2 \times 10^{32}$ years
 for $p \rightarrow  e^+ \nu \nu$ and $p \rightarrow \mu^+ \nu \nu$, respectively.
The sensitivity to these modes 
is calculated to be $2.7 \times 10^{32}$ and $2.5 \times 10^{32}$ years.
The lifetime limits found in this study are an order of magnitude improvement over the
 previous results \cite{McGrew:1999nd, Berger:1991fa}.
These results provide strong constraints to both 
the permitted parameter space of Refs. \cite{Pati:1983jk, Gu:2011pf},
 which predict lifetimes of around
$10^{30} - 10^{33}$ years, 
and on other GUT models which allow for similar processes.
\VT{We note, that the analyses presented in this work are only weakly model dependent,
due to the assumption of a flat phase space in the signal generation.
However, this assumption agrees well with alternative phase space considerations \cite{Chen:2014ifa}
in the context of vector- or scalar-mediated proton decays,
which are typical of GUT models  \cite{Georgi:1974sy,Pati:1974yy, Fritzsch:1974nn}.}

~\newline
\indent 
We gratefully acknowledge cooperation of the Kamioka Mining and Smelting
Company. The Super-Kamiokande experiment was built and has been operated with
funding from the Japanese Ministry of Education, Culture, Sports, Science and Technology, the United States Department of Energy, and the U.S. National Science Foundation. Some of us
have been supported by funds from the Korean Research Foundation (BK21), the
National Research Foundation of Korea (NRF-20110024009), the State Committee for 
Scientific Research in Poland (grant1757/B/H03/2008/35), the Japan Society for
Promotion of Science, and the National Natural Science Foundation of China under
Grants No.10575056.
%

\bibliography{nucleonbib}

\end{document}